\documentclass[a4paper,aps,twocolumn,pre,intlimits]{revtex4-1}
\usepackage{amsmath,amsthm,amssymb,amsthm,dsfont,mathrsfs,upgreek}
\usepackage{mathtools,xfrac}
\usepackage[font=small,justification=centerlast,format=plain]{caption}
\usepackage[font=small,justification=centerlast,format=plain]{subcaption}
\usepackage{tikz}
\usepackage[english]{babel}

\newcommand{\R}{\mathds{R}}

\newcommand{\Q}{\mathds{Q}}
\newcommand{\C}{\mathds{C}}

\DeclareMathOperator{\dd}{d}
\DeclareMathOperator{\e}{e}

\usepackage{hyperref}
\hypersetup{pdfauthor={Sicuro Tsallis},pdftitle={qDirac},%
            colorlinks, linktocpage=true, pdfstartpage=3, pdfstartview=FitV,%
    breaklinks=true, pdfpagemode=UseNone, pageanchor=true, pdfpagemode=UseOutlines,%
    plainpages=false, bookmarksnumbered, bookmarksopen=true, bookmarksopenlevel=1,%
    hypertexnames=true, pdfhighlight=/O,%
    urlcolor=blue, linkcolor=blue, citecolor=red, %pagecolor=RoyalBlue,%
        }

\usepackage[T1]{fontenc}
\usepackage{lmodern}
\usepackage[cal=boondoxo]{mathalfa}
\allowdisplaybreaks
\begin{document}
%\title{$q$-Generalization of $d$ dimensional Dirac delta and Fourier transform, and of convolution product}
\title{$q$-Generalized representation of the $d$-dimensional Dirac delta and $q$-Fourier transform}
\author{Gabriele Sicuro}\email{sicuro@cbpf.br}
\affiliation{Centro Brasileiro de Pesquisas F\'isicas, Rua Dr. Xavier Sigaud, 150, 22290-180, Rio de Janeiro, Brazil}
\author{Constantino Tsallis}\email{tsallis@cbpf.br}
\affiliation{Centro Brasileiro de Pesquisas F\'isicas, and National Institute of Science and Technology of Complex Systems, Rua Xavier Sigaud 150, 22290-180 Rio de Janeiro-RJ,  Brazil}
\affiliation{Santa Fe Institute, 1399 Hyde Park Road, New Mexico 87501, USA}
\affiliation{Complexity Science Hub Vienna, Josefst\"adter Strasse 39, 1080 Vienna, Austria}

\date{\today}
\begin{abstract}We introduce a generalized representation of the Dirac delta function in $d$ dimensions in terms of $q$-exponential functions. We apply this new representation to the study of the so-called $q$-Fourier transform and we establish the analytical procedure through which it can be inverted for any value of $d$. We finally illustrate the effect of the $q$-deformation on the Gibbs phenomenon of Fourier series expansions.
\end{abstract}
\maketitle

\paragraph{Introduction}\label{intro}
Since the foundational works of Clausius, Boltzmann and Gibbs, entropy has had a central role in the theory of thermodynamics and statistical mechanics \cite{Callen1985,Greven2014,Lieb1999}. In the classical Boltzmann--Gibbs (BG) theory of statistical mechanics, the entropy associated to a certain macrostate is expressed in terms of a probability distribution $\mathbf p=\{p_i\}_{i=1,\dots, W}$ given on the set of $W\in\mathds N$ microstates that are compatible with the considered macrostate, through the relation
\begin{equation}\label{BGen}
S_{\text{BG}}[\mathbf p]\coloneqq -\sum_{i=1}^W p_{i}\ln p_i
\end{equation}
(we have fixed the Boltzmann constant equal to one). The expression above, that is valid for classical mixing/ergodic systems, provides a connection between the macroscopic and the microscopic description of the system. Shannon \cite{Shannon1948}, and later Khinchin \cite{Khinchin1957}, showed that the concept of entropy, and its expression in Eq.~\eqref{BGen}, play also a fundamental role in information theory, and it is indeed at the heart of the theory of communication. Due to the relevance of the concept of entropy both in physics and in information theory, during the last decades many generalizations of the BG entropy have been proposed \cite{Beck1995,Tempesta2015,Enciso2017,Tsallis2009B}. In particular, the R\'enyi entropy \cite{Renyi1961}
\begin{subequations}\label{genen}
\begin{equation}
S_q^{\text{R}}[\mathbf p]=\frac{\ln\left(\sum_{i=1}^Wp_i^q\right)}{1-q},\quad q \in\R \,,
\end{equation}
and the nonadditive entropy $S_q$ \cite{Tsallis1988}
\begin{equation}
S_q[\mathbf p]=\frac{1-\sum_{i=1}^W p_i^q}{q-1},\quad q \in\R\,.
\end{equation}\end{subequations}
Both entropies generalize the BG one, which is recovered for $q\to 1$. The 
nonadditive entropy $S_q$ has been profusely investigated and applied to the 
study of a wide spectrum of physical properties of complex systems 
\cite{Beck1995,Tsallis2009B}.  The parameter $q$ measures the deviation from the 
classical case. In the spirit of the maximum entropy principle introduced by 
Jaynes \cite{Jaynes1957}, it can be shown that the generalized entropies in 
Eqs.~\eqref{genen} are extremized by the same family of distributions under the 
same kind of constraints \cite{Lenzi2000,Tsallis1998}. Indeed, denoting by 
$\{\epsilon_i\}_{ i = 1 ,\dots, W}$ the energies corresponding to the $W$ 
microstates appearing in the sums in Eqs.~\eqref{genen}, imposing a fixed 
average energy constraint\footnote{Observe that, under the assumption that the 
value of $\beta$ is kept fixed, the constraint must be imposed using escort 
distributions (see below) to guarantee the invariance of the maximizing 
distribution under energy shifts. See also Refs.~\cite{Lenzi2000,Tsallis1998} 
for additional details.} as
\begin{equation}
\frac{\sum_{i=1}^W p_i^q\epsilon_i}{\sum_{i=1}^W p_i^q}=\langle\epsilon\rangle_q,
\end{equation}
the maximizing distribution of the entropies in Eqs.~\eqref{genen} is
\begin{equation}\label{qexpE}
p_i(\epsilon_i)=\frac{1}{Z}\exp_q\left[-\beta\left(\epsilon_i-\langle\epsilon\rangle_q\right)\right].
\end{equation}
where $Z$ is a proper normalization factor, $\beta>0$ is a Lagrange multiplier and, for $q\in\R$,
\begin{multline}
\exp_q(x)\coloneqq \left[1+(1-q)x\right]^\frac{1}{1-q}_+,\quad 
[x]_+\coloneqq x\theta(x)\label{qexpcl}
\end{multline}
is the so called \textit{$q$-exponential function}. The $q$-exponential function generalizes the usual exponential, which is recovered in the $q\to 1$ limit. 

Probability distributions, as well as other physical quantities, having a 
$q$-exponential shape are found in the analysis of data obtained in high-energy 
experiments \cite{HEP2010}, finance \cite{Borland2002}, dusty plasmas 
\cite{Liu2008}, and theoretical investigations on optical lattices 
\cite{Renzoni2006,Lutz2013}, low-dimensional dissipative maps 
\cite{Ananos2004,Robledo2004,Lyra1998}, diffusion processes in superconductors 
\cite{Andrade2010}, among many others. The ubiquity of distributions with 
power-law tails in the form of Eq.~\eqref{qexpE} has suggested the existence of 
a $q$-generalized Central Limit Theorem (CLT) for some classes of correlated 
random variables, in analogy with the connection between Maxwell distribution in 
BG statistical mechanics and the usual CLT for uncorrelated (or weakly 
correlated) random variables. This possibility has been investigated, for 
example, by the authors of Ref.~\cite{uts2008} and led, as by-product, to a 
generalization of many standard mathematical concepts \cite{Jauregui2010}, on 
the basis of a new deformed algebra previously introduced by Borges 
\cite{Borges2004}. In particular, a possible generalization of the usual Fourier 
transform, called \textit{$q$-Fourier transform} ($q$-\textsc{FT}) was proposed 
in Ref.~\cite{uts2008}. The new integral transform was defined in formal analogy 
with the usual Fourier transform and expressed in terms of the analytic 
prolongation of the deformed exponential function given in Eq.~\eqref{qexpcl}. 
Hilhorst \cite{Hilhorst2010} observed however that the $q$-\textsc{FT} as 
defined in Ref.~\cite{uts2008} cannot be inverted. For this reason, a modified 
definition of $q$-\textsc{FT}, that is invertible, has been proposed in 
Ref.~\cite{Jauregui2011}.

Inspired by the results discussed above, in the present paper we analyze a generalization of the $q$-\textsc{FT}, in the form adopted in Ref.~\cite{Jauregui2011}, to the $d$-dimensional case. The invertibility of this expression is proven using a new representation of the Dirac delta function in $d$ dimensions, based again on $q$-exponentials. We will finally give some numerical examples, discussing a series representation of the inverse $q$-\textsc{FT} and comparing it with the classical Fourier series.

\begin{figure}\centering
\includegraphics[width=\columnwidth]{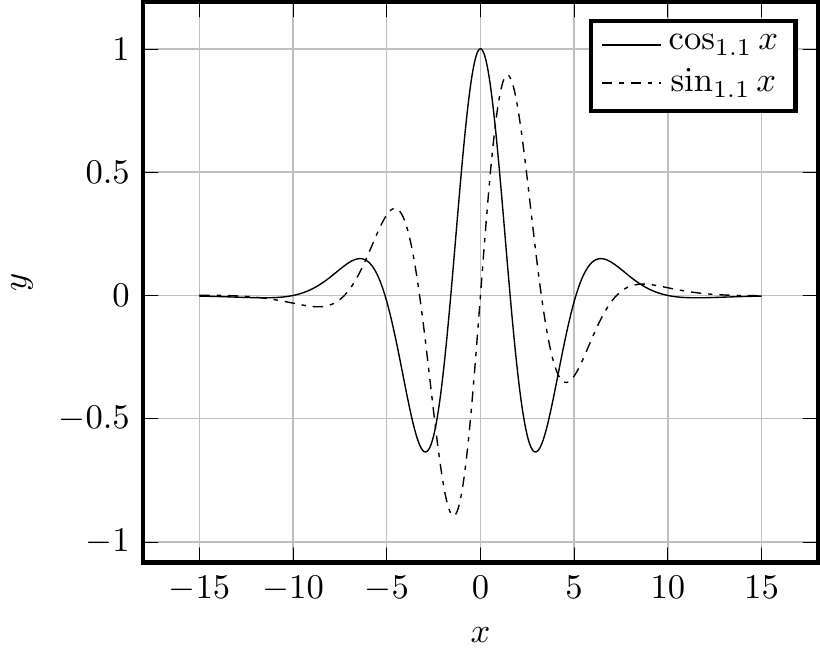}
\caption{Real and imaginary part of the $q$-plane wave $\e_q^{ix}$ for $q=1+\sfrac{1}{10}$.}\label{f:qexpw}
\end{figure}

\paragraph{Preliminaries: the $q$-exponential}
\begin{figure}\centering
\includegraphics[width=\columnwidth]{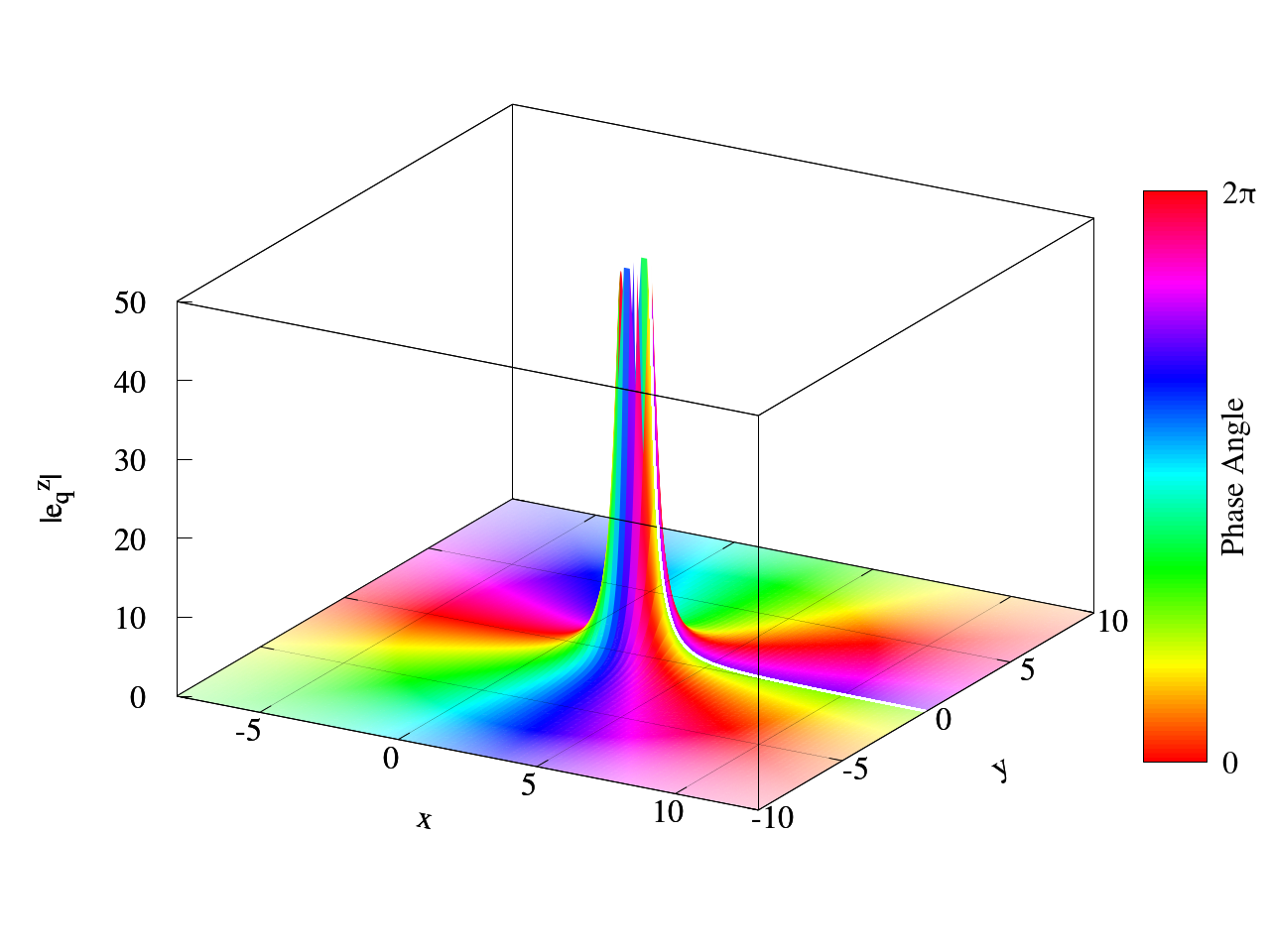}
\caption{Plot of the modulus $|\e_q^z|$ of the $q$-exponential function in Eq.~\eqref{qexp} for $q=\sfrac{7}{5}$ on the complex plane respect to the variable $z=x+iy$. In color, the complex argument. Observe the presence of the pole at $z_q=\sfrac{5}{2}$. The branch cut in the complex plane corresponds to the half real line $[\sfrac{5}{2},+\infty)$.}\label{f:qexp}
\end{figure}
In the present paper we will use the \textit{$q$-exponential function} defined as
\begin{equation}
\e_{q}^z\colon\C\to\C,\quad \e_{q}^z\coloneqq \left[1+(1-q)z\right]^{\frac{1}{1-q}},\quad q\in\R.\label{qexp}
\end{equation}
The previous definition is not, strictly speaking, the usual one adopted in the literature, given in Eq.~\eqref{qexpcl}, but instead its analytic continuation to the complex plane. Indeed, the $q$-exponential in Eq.~\eqref{qexpcl} is a real function of a real variable, and a cutoff appears in Eq.~\eqref{qexpcl} that is absent in Eq.~\eqref{qexp}. If $z\in\R$, the two definitions coincide for $z\lessgtr z_q$ if $q\gtrless 1$, where
\begin{equation}\label{cov}
z_q=\frac{1}{q-1}.
\end{equation}
In the following, we will consider $q> 1$ only. For $q>1$ the analytic properties of the $q$-exponential are analogous to the ones of the function $f(z)=z^{-a}$, $a>0$, on the complex plane. In particular, the $q$-exponential has a pole for $z=z_q$. We choose as branch cut the half line $[z_q,+\infty)$ along the positive real axis, see Fig.~\ref{f:qexp}. If $z_q\in\Q^+$, we can construct a Riemann surface for the $q$-exponential with a finite number of branches. If otherwise $z_q\in\R^+\setminus\Q$, the number of branches is infinite. Analyticity is recovered for $z_q\to+\infty$, i.e., for $q\to 1$, when
\begin{equation}\label{limite}
\lim_{q\to 1^\pm}\e_{q}^z=\e^z.
\end{equation}
The $q$-exponential function is therefore a deformation of the usual exponential. Many properties of the usual exponential are, however, lost for $q\neq 1$. For example, for $q\neq 1$ and $z,w\in\C$,
\begin{equation}
\e_{q}^{z+w}\neq \e_{q}^{z}\e_{q}^{w}\text{ and }\e_{q}^{z}\neq\e_{q}^{z+2\pi i}.
\end{equation}
The $q$-exponential function in Eq.~\eqref{qexp} can be also written, for $q>1$ and $\Re(z)<z_q$, as a linear superposition of exponentials, in the form \cite{Jeffrey2007}
\begin{multline}\label{super}
\e_{q}^z=\int_0^{+\infty}\gamma\left(z_q;t\right)\e^{tz}\dd t,\quad \Re(z)<z_q,\\ \gamma(\eta;t)\coloneqq\frac{\eta^\eta t^{\eta-1}}{\Gamma(\eta)}\e^{-\eta t},\quad \eta>0.
\end{multline}
Observe that, as expected,
\begin{equation}
\lim_{q\to 1^+}\gamma\left(z_q;t\right)=\delta(t-1).
\end{equation}
Eq.~\eqref{super} has been extensively discussed by Beck \cite{Beck2003} in the context of superstatistics. Eq.~\eqref{super} implies that, given $x\in\R$ and $q>1$,
\begin{equation}
\e_{q}^{ix}=\int_0^{+\infty}\gamma\left(z_q;t\right)\e^{itx}\dd t.\label{super2}
\end{equation}
The function $\e_q^{ikx}$, with $x\in\R$ and $k\in\R$, is sometimes called $q$-plane wave of momentum $k$ \cite{Jauregui2010}. Its real part and imaginary part are a deformation of the cosine function and the sine function respectively. In particular, denoting by
\begin{subequations}\label{qtrig}
\begin{equation}
\omega_n(q)\coloneqq\prod_{k=0}^n\left[k(q-1)+1\right]\xrightarrow{q\to 1}1,
\end{equation}
Borges \cite{Borges2004} introduced the generalized $q$-cosine and $q$-sine functions (see Fig.~\ref{f:qexpw}) as
\begin{equation}
\e_q^{\pm ix}=1+\sum_{n=1}^\infty \omega_{n-1}(q)\frac{(\pm i x)^n}{n!}\equiv\cos_q x\pm i\sin_q x,\end{equation}
where
\begin{align}
\cos_q x&\coloneqq 1+\sum_{n=1}^\infty \omega_{2n-1}(q)\frac{(-1)^n x^{2n}}{(2n)!},\\
\sin_q x&\coloneqq \sum_{n=0}^\infty \omega_{2n}(q)\frac{(-1)^n x^{2n+1}}{(2n+1)!}.
\end{align}
\end{subequations}

\paragraph{A representation of the Dirac delta function}\label{dirac} 
In Ref.~\cite{Jauregui2010} a representation of the Dirac delta function on the real line in terms of $q$-exponentials has been proposed. The proof of this new representation has been obtained by different authors \cite{Chevreuil2010,Mamode2010,Plastino2011} using different approaches. In the following, we will generalize these results, showing that an integral representation for the Dirac delta function on $\R^d$ can be obtained using again $q$-exponentials. Let us introduce the following distribution
\begin{equation}
\delta_{q}^{(d)}(\mathbf x)\coloneqq\frac{1}{c(q,d)}\int_{\R^d}\e_{q}^{i\mathbf k\cdot\mathbf x}\dd^dk,\quad q\in\left(1,1+\frac{1}{d}\right).\label{deltag}
\end{equation}
In the previous expression, $c(q,d)$ is a numerical prefactor depending on $q$ and $d$ that we will suitably fix later on. We will show now that the function $\delta_{q}^{(d)}(\mathbf x)$ behaves like a Dirac delta function in $d$ dimensions. Indeed, given a test function $\varphi(\mathbf x)$ that is infinitely differentiable and rapidly decreasing at infinity, we have that
\begin{multline}
\int_{\R^d}\varphi(\mathbf x)\delta^{(d)}_q\left(\mathbf x\right)\dd^dx=\\=\frac{1}{c(q,d)}\int_{\R^d}\left[\int_{\R^d}\varphi(\mathbf x)\e_{q}^{i\mathbf k\cdot\mathbf x}\dd^dx\right]\dd^dk.
\end{multline}
Using Eq.~\eqref{super2} we can write
\begin{multline}
\int_{\R^d}\varphi(\mathbf x)\e_{q}^{i\mathbf k\cdot\mathbf x}\dd^dx
=\int_0^{+\infty}\left[\int_{\R^d}\varphi(\mathbf x)\e^{it\mathbf k\cdot\mathbf x}\dd^dx\right]\gamma(z_q;t)\dd t\\
=\int_0^{+\infty}\hat\varphi(t\mathbf k)\gamma(z_q;t)\dd t,
\end{multline}
where
\begin{equation}
\hat\varphi(\mathbf k)\coloneqq \int_{\R^d}\varphi(\mathbf x)\e^{i\mathbf k\cdot\mathbf x}\dd^dx
\end{equation}
is the Fourier transform of $\varphi$. Using the fact that 
$\int_{\R^d}\hat\varphi(\mathbf k)\dd^dk=\varphi(\mathbf 0)$, we finally have
\begin{equation}
\int_{\R^d}\varphi(\mathbf x)\delta^{(d)}_{q}\left(\mathbf x\right)\dd^dx=\left(\frac{2\pi}{q-1}\right)^d\frac{\Gamma\left(z_q-d\right)}{c(q,d)\Gamma\left(z_q\right)}\varphi(\mathbf 0),
\end{equation}
under the condition
\begin{equation}
1<q<1+\frac{1}{d}\,.
\end{equation}
If we impose now
\begin{equation}
c(q,d)\coloneqq\left(\frac{2\pi}{q-1}\right)^d\frac{\Gamma\left(z_q-d\right)}{\Gamma\left(z_q\right)},\quad z_q\coloneqq\frac{1}{q-1},
\end{equation}
we obtain the result
\begin{equation}
\int_{\R^d}\varphi(\mathbf x)\delta^{(d)}_{q}\left(\mathbf x\right)\dd^dx=\varphi(\mathbf 0),\quad 1<q<1+\frac{1}{d},
\end{equation}
i.e., the function $\delta_{q}^{(d)}$ acts as a Dirac distribution function on $\R^d$. Observe that, as expected,
\begin{equation}
\lim_{q\to 1^+}c\left(q,d\right)=(2\pi)^d\Rightarrow\lim_{q\to 1^+}\delta^{(d)}_{q}\left(\mathbf x\right)=\delta^{(d)}\left(\mathbf x\right).
\end{equation}
The results above recover the $d=1$ case analyzed in Refs.~\cite{Chevreuil2010,Jauregui2010,Mamode2010,Plastino2011} as particular case.

\paragraph{The $q$-Fourier transform in $d$ dimensions, and its inversion}
The interest of the authors of Ref.~\cite{Jauregui2010} in the function $\delta_{q}^{(1)}(x)$ was motivated by the problem of the inversion of the so called \textit{$q$-Fourier transform} ($q$-\textsc{FT}). The $q$-\textsc{FT} of a real, nonnegative integrable function $f(x)$ is given by \cite{uts2008}
\begin{multline}\label{qFT1}
\hat f_{q}(k)\coloneqq \int_{-\infty}^{+\infty}f(x)\odot_{q}\e_{q}^{ikx}\dd x\\
\equiv \int_{-\infty}^{+\infty}f(x)\e_{q}^{ikx\left[f(x)\right]^{q-1}}\dd x,\quad 1<q<2.
\end{multline}
In the previous expression, we have introduced the binary operation $\odot_{q}$ between two complex numbers $z,w\in\C$, defined as
\begin{equation}\label{qprod}
z\odot_{q} w\coloneqq \left(z^{1-q}+w^{1-q}-1\right)^\frac{1}{1-q},\quad q>1.
\end{equation}
Observe that the operation above is a generalization of the usual product, being
\begin{equation}
z\odot_{q} w\xrightarrow{q\to 1^+} zw.
\end{equation}
The operation $\odot_q$ is abelian, $z\odot_{q} w=w\odot_{q} z$, and such that $z\odot_{q} 1=z$ and $\lim_{w\to 0}z\odot_{\eta} w=0$ $\forall z\neq 0$. Moreover, the operation introduced in Eq.~\eqref{qprod} generalizes the so-called \textit{$q$-product} between two real positive quantities, as defined in Ref.~\cite{Borges2004}, namely
\begin{equation}\label{qprodcl}
x\otimes_q y\coloneqq \left[x^{1-q}+y^{1-q}-1\right]_+^\frac{1}{1-q},\quad x,y>0.
\end{equation}
Eq.~\eqref{qprodcl} coincides with Eq.~\eqref{qprod} for $z,w\in\R^+$ and $z^{1-q}+w^{1-q}\geq 1$. The $q$-\textsc{FT} defined in Eq.~\eqref{qFT1} recovers therefore the usual Fourier transform for $q\to 1^+$. However, the expression in Eq.~\eqref{qFT1} requires some further discussion. Indeed, as first noted by Hilhorst \cite{Hilhorst2010}, the introduced integral transform cannot be, in general, inverted. This observation affects possible applications of Eq.~\eqref{qFT1}, and, for this reason, an extended definition of $q$-\textsc{FT} has been introduced in Ref.~\cite{Jauregui2011}, namely
\begin{multline}\label{qFT2}
\hat f_{q}(k;kx)\coloneqq \int_{-\infty}^{+\infty}f(y)\odot_{q}\e_{q}^{ik(y-x)}\dd y
\\
\equiv \int_{-\infty}^{-\infty}f(y)\e_q^{ik(y-x)\left[f(y)\right]^{q-1}}\dd y\xrightarrow{q\to 1^+}\e^{-ikx}\hat f(k).
\end{multline}
The introduction of the shift variable $x$ allows us to invert the integral transform for $1<q<2$. Observe that the variable $x$ appears in the product $kx$ only: we have stressed this fact in our notation. It can be proved that \cite{Jauregui2011}
\begin{equation}\label{invft1}
f(x)=\left[\frac{2-q}{2\pi}\int_{-\infty}^{+\infty} \hat f_{q}(k;kx)\dd k\right]^{\frac{1}{2-q}},\quad q\in(1,2).
\end{equation}
The proof of Eq.~\eqref{invft1} strongly relies on the properties of the function $\delta_{q}^{(1)}(x)$ discussed above. Moreover, the standard formula for the inversion of the Fourier transform is obtained for $q\to 1^+$.

We present here a derivation of the result in Eq.~\eqref{invft1} in a more general setting, namely considering a generalization of Eq.~\eqref{qFT2} to $d$ dimensions. Given a nonnegative function $f\colon\R^d\to\R^+$, we define its $q$-\textsc{FT} in $d$ dimensions as
\begin{multline}\label{qFTd}
\hat f_{q}(\mathbf k;\mathbf k\cdot \mathbf x)\coloneqq\int_{\R^d} f(\mathbf y)\odot_{q}\e_{q}^{i\mathbf k\cdot(\mathbf y-\mathbf x)}\dd^dy\\
\equiv \int_{\R^d} f(\mathbf y)\e_{q}^{i\mathbf k\cdot(\mathbf y-\mathbf x)\left[f(\mathbf y)\right]^{q-1}}\dd^dy,\\\text{for } q\in\left(1,1+\frac{1}{d}\right).
\end{multline}
The previous expression straightforwardly generalizes Eq.~\eqref{qFT2}, apart from the nontrivial constraint $q\in\left(1,1+\frac{1}{d}\right)$, which guarantees invertibility. The inversion of the integral transform in Eq.~\eqref{qFTd} can be obtained using the function $\delta_{q}^{(d)}$ discussed in the previous section. Indeed, assuming that $q\in\left(1,1+\frac{1}{d}\right)$,
\begin{multline}
\int_{\R_d}\hat f_{q}(\mathbf k;\mathbf k\cdot \mathbf x)\dd^dk\\
=\int_{\R^d} f(\mathbf y)\left[\int_{\R_d}\e_{q}^{i\mathbf k\cdot(\mathbf y-\mathbf x)\left[f(\mathbf y)\right]^{q-1}}\dd^d k\right]\dd^dy\\
=c(q,d)\int_{\R^d} f(\mathbf y)\delta_{q}^{(d)}\left((\mathbf y-\mathbf x)\left[f(\mathbf y)\right]^{q-1}\right)\dd^dy\\
=c(q,d)\int_{\R^d}\left[f(\mathbf y)\right]^{1-d(q-1)}\delta^{(d)}\left(\mathbf y-\mathbf x\right)\dd^dy\\
=c(q,d)\left[f(\mathbf x)\right]^{1-d(q-1)}.
\end{multline}
In the last step we have supposed that $\mathbf x$ belongs to the interior of the support of $f$ (otherwise an additional factor will appear, due to the fact that the Dirac delta function is evaluated on a boundary point \cite{Jauregui2011}). It follows that
\begin{equation}
f(\mathbf x)=\left[\frac{1}{c(q,d)}\int_{\R^d} \hat f_{q}(\mathbf k;\mathbf k\cdot \mathbf x)\dd^dk\right]^{\frac{1}{1-d(q-1)}}.\label{qFTdinv}
\end{equation} 
Eq.~\eqref{qFTdinv} generalizes the result in Eq.~\eqref{invft1}, which is indeed recovered for $d=1$. Moreover, the standard relation between $f$ and its Fourier transform is obtained for $q\to 1^+$.% Finally, observe also that $d\to+\infty$ implies that our formulas are valid for $q\to 1^+$ only.
\paragraph{On series expansion and $q$-\textsc{FT}}
Let us now consider a positive real function 
$f\colon[-\sfrac{T}{2},\sfrac{T}{2}]\to\R^+$ that is square integrable on its 
domain. It is well known that $f$ can be represented in terms of a Fourier 
series, that in the notation introduced in Eq.~\eqref{qFTd} reads
\begin{multline}\label{Fseries}
f(x)%=\sum_{n=-\infty}^{+\infty}f_n \e^{-\frac{2\pi inx}{T}},\quad f_n\coloneqq \frac{1}{T}\int_{-\frac{T}{2}}^{\frac{T}{2}}f(x)\e^{\frac{2\pi inx}{T}}\dd x\\
=\frac{1}{T}\sum_{n=-\infty}^{+\infty}\hat f_1\left(\frac{2\pi n}{T};\frac{2\pi n}{T}x\right)\\
=\frac{1}{T}\sum_{n=-\infty}^{+\infty}\e^{-\frac{2\pi i n}{T}x}\hat f\left(\frac{2\pi n}{T}\right).
\end{multline}
In the $T\to\infty$ limit we formally recover the expression of the Fourier transform and its inverse. 

It is tempting to generalize the previous standard result deforming the circular 
functions according to Eqs.~\eqref{qtrig}. However, it is easily verified 
that the orthogonality condition among the $q$-deformed functions in 
Eq.~\eqref{qtrig} is not satisfied, and therefore the quantities in 
Eqs.~\eqref{qtrig} cannot be considered a basis in a Hilbert space. Despite this 
important fact, for $q\in(1,2)$, we have that, for a given positive function $f$ 
on the real line,
\begin{multline}
f^{(q)}(x)\coloneqq\left[\frac{2-q}{T}\sum_{n=-\infty}^{+\infty}\hat f_q\left(\frac{2\pi n}{T};\frac{2\pi n}{T}x\right)\right]^{\frac{1}{2-q}}\\
\xrightarrow{T\to\infty}\left[\frac{2-q}{2\pi}\int_{-\infty}^{+\infty} \hat f_{q}(k;kx)\dd k\right]^{\frac{1}{2-q}}=f(x) \,.
\end{multline}
If we consider therefore a function $f$ with compact support in $[-\sfrac{T}{2},\sfrac{T}{2}]$, it is expected that, although $f^{(q)}\neq f$ for $q\neq 1$, the sum
\begin{equation}
S_N^{(q)}[f](x)\coloneqq \left[\frac{2-q}{T}\sum_{n=-N}^{N}\hat f_q\left(\frac{2\pi n}{T};\frac{2\pi n}{T}x\right)\right]^{\frac{1}{2-q}},\label{sN}
\end{equation}
provides a good approximation of $f$ for $N\gg1$ and, if $q\neq 1$, for $T\gg 1$. In particular, we expect that for $q\approx 1$, the condition $T\gg 1$ can be relaxed and the expression $S_N^{(q)}[f]$ still provides a good approximation of $f$ for $N\gg 1$, despite the fact that no specific periodicity can be associated to a $q$-plane wave for $q\neq 1$. 

\begin{figure*}[htbp]
\begin{subfigure}[t]{0.45\textwidth}
 \includegraphics[width=\textwidth]{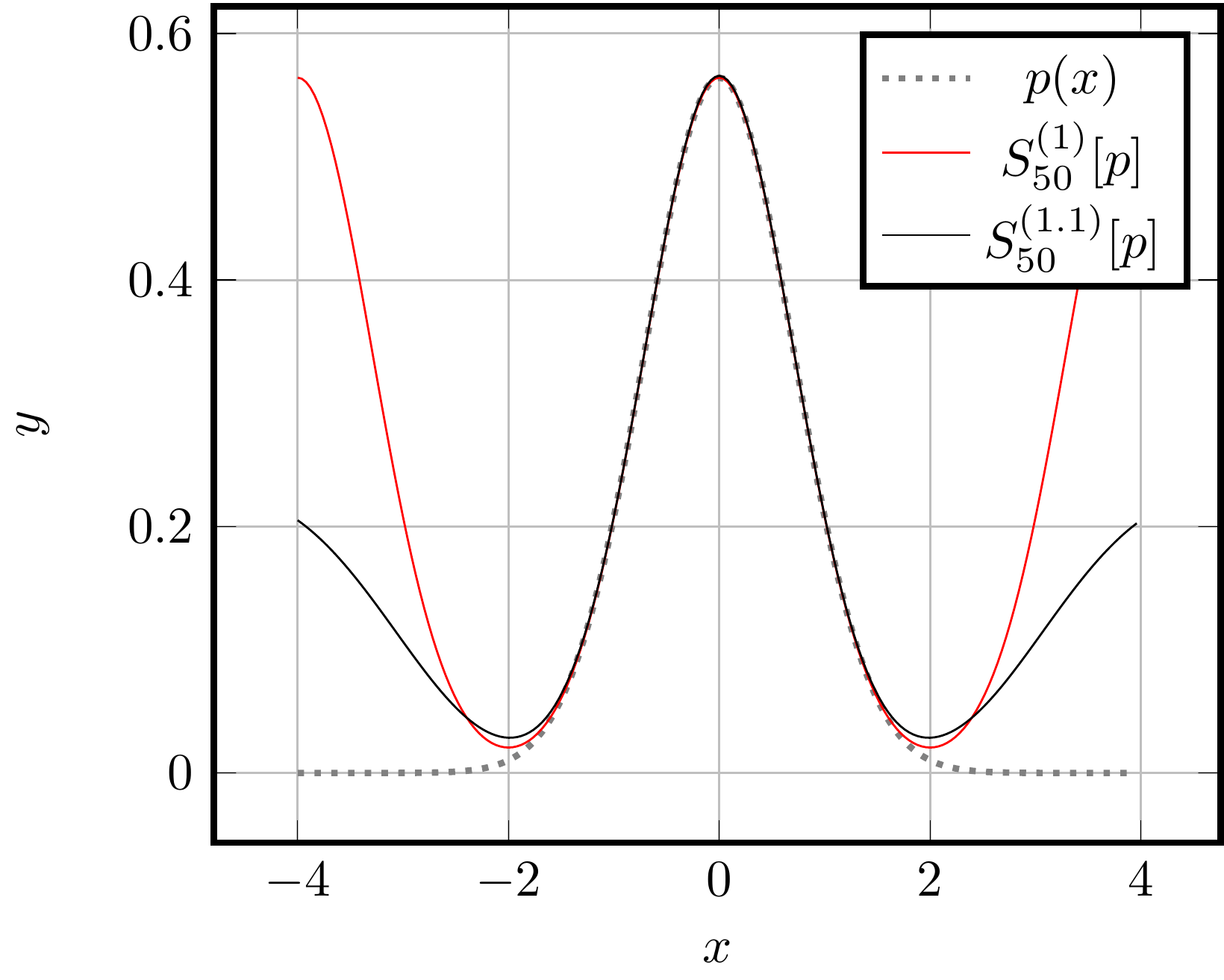}
 \caption{Approximation of the function in Eq.~\eqref{gaussiana} obtained using Eq.~\eqref{sN} with $N=50$ and for different values of $q$.\label{f:gibbsg}}
\end{subfigure}\hspace{1cm}
\begin{subfigure}[t]{0.45\textwidth}
 \includegraphics[width=\textwidth]{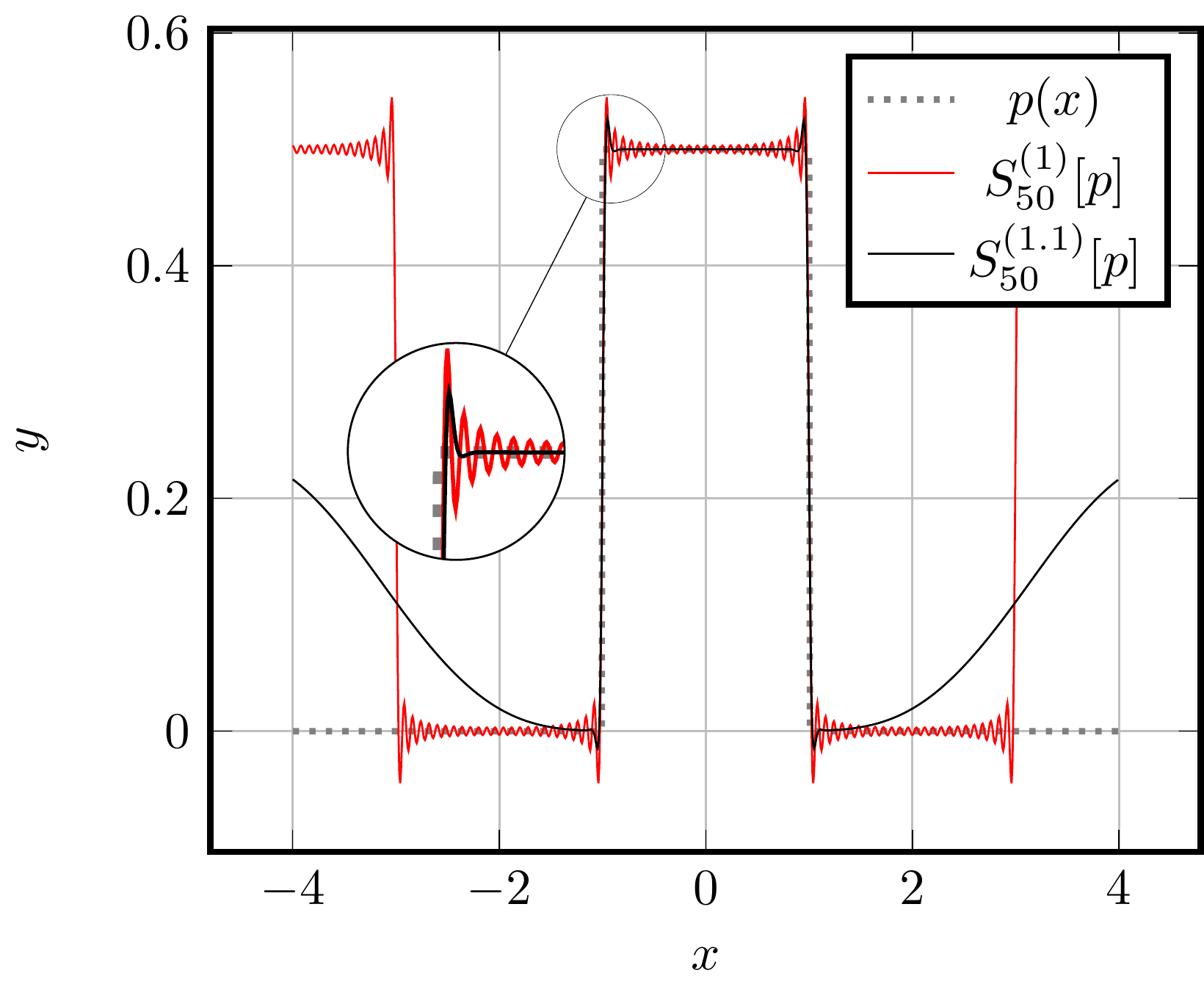}
 \caption{Approximation of the function in Eq.~\eqref{square} obtained using Eq.~\eqref{sN} with $N=50$ and for different values of $q$. Observe that, for say $q=1.1$, the Gibbs phenomenon is sensibly attenuated.\label{f:gibbs}}
\end{subfigure}
\caption{Examples of application of Eq.~\eqref{sN} to different probability distribution densities.}
\end{figure*}

To numerically illustrate this fact, let us consider, for example, $q=1+\sfrac{1}{10}$, and two different distribution densities. Let us first start with a smooth distribution, namely a Gaussian distribution
\begin{equation}\label{gaussiana}
 p(x)=\frac{\e^{-x^2}}{\sqrt{\pi}},
\end{equation}
and let us apply Eq.~\eqref{sN} to it on the interval $[-2,2]$. In Fig.~\ref{f:gibbsg} we show that both $S_{50}^{(1)}[p]$ and $S_{50}^{(1.1)}[p]$ approximate very well the function $p$ on the considered domain.

Let us now consider the uniform distribution
\begin{equation}\label{square}
p(x)=2\frac{\theta\left(\frac{T}{4}-x\right)\theta\left(x+\frac{T}{4}\right)}{T}
\end{equation}
on the interval $[-\sfrac{T}{2},\sfrac{T}{2}]$. In Fig.~\ref{f:gibbs} we show the approximation for $p(x)$ in the case $T=4$ obtained using $S_N^{(1)}[p]$ and $S_N^{(1.1)}[p]$ for $N=50$. Remarkably, the approximation provided by the expansion obtained for $q=1.1$ does not fluctuate in the interval $[-\sfrac{T}{4},\sfrac{T}{4}]$, where $p$ is different from zero and the Gibbs phenomenon appears to be suppressed, even close to the discontinuity points, despite the fact that both series have been truncated at the same value of $N$. The possible application of the series in Eq.~\eqref{sN}, and the effects of the $q$-deformation on the Gibbs phenomenon, still deserve further investigation.

\paragraph{Conclusions}
In the present paper, we have discussed a new representation of the Dirac delta function in $d$ dimensions, given in terms of $q$-exponential functions. Using this representation, we have proved the invertibility of the $q$-\textsc{FT} in $d$ dimensions. We have finally numerically illustrated the effects of the $q$-deformation on the Gibbs phenomenon in a Fourier-like series expansion.

The new tools introduced in the present paper are expected to be useful in further investigations on the mathematical foundations of generalized thermostatistics. In particular, the $q$-\textsc{FT} in $d$ dimensions and the $q$-generalized representation of the Dirac delta function can be useful in the study of generalized versions of the Central Limit Theorem. A more detailed and rigorous investigation of the applicability of the previous tools is also of great mathematical interest.

\paragraph*{Acknowledgments}
The authors thank Max J\'auregui for useful discussions. They also acknowledge partial financial support by CNPq and Faperj (Brazilian agencies) and by the John Templeton Foundation (USA).

\end{document}